\documentclass[fdp, fleqn]{w-art}

\usepackage{subfigure}
\usepackage{cite}
\usepackage{times}
\usepackage[hyperref,bulletsep]{collect}
\usepackage{bbm}
\usepackage{amsmath}
\usepackage{amssymb}
\usepackage{amscd}
\usepackage{slashed}
\usepackage{array}
\usepackage{rotating}
\usepackage{float}
\usepackage{subfigure}
\usepackage{feynmp}

\makeatletter \@addtoreset{equation}{section} \makeatother

\makeatletter
\let\old@startsection=\@startsection
\let\oldl@section=\l@section
\renewcommand{\@startsection}[6]{\old@startsection{#1}{#2}{#3}{#4}{#5}{#6\mathversion{bold}}}
\renewcommand{\l@section}[2]{\oldl@section{\mathversion{bold}#1}{#2}}
\makeatother

\makeatletter
\let\old@makecaption=\@makecaption
\def\@makecaption{\small\old@makecaption}
\makeatother

\renewcommand{\thefootnote}{\arabic{footnote}}
\let\oldPhi=\Phi
\let\oldPsi=\Psi
\let\oldGamma=\Gamma
\let\oldDelta=\Delta
\let\oldSigma=\Sigma
\let\oldTheta=\Theta
\let\oldPi=\Pi
\let\oldUpsilon=\Upsilon
\renewcommand{\Phi}{\mathnormal{\oldPhi}}
\renewcommand{\Psi}{\mathnormal{\oldPsi}}
\renewcommand{\Gamma}{\mathnormal{\oldGamma}}
\renewcommand{\Sigma}{\mathnormal{\oldSigma}}
\renewcommand{\Delta}{\mathnormal{\oldDelta}}
\renewcommand{\Theta}{\mathnormal{\oldTheta}}
\renewcommand{\Pi}{\mathnormal{\oldPi}}
\renewcommand{\Upsilon}{\mathnormal{\oldUpsilon}}

\newcommand{\tr}{\mathop{\mathrm{tr}}}

\ifx\genfrac\sdflkaj

\else

\fi

\newcommand{\bigbrk}[1]{\bigl(#1\bigr)}

\newcommand{\Bigsbrk}[1]{\Bigl[#1\Bigr]}

\newcommand{\nl}[1][0pt]{\nonumber\\[#1]&\hspace{-4\arraycolsep}&\mathord{}}

\newcommand{\earel}[1]{\mathrel{}&\hspace{-2\arraycolsep}#1\hspace{-2\arraycolsep}&\mathrel{}}
\newcommand{\eq}{\earel{=}}

\def\[{\begin{equation}}
\def\]{\end{equation}}

\newcommand{\be}{\begin{eqnarray}}
\newcommand{\ee}{\end{eqnarray}}

\makeatletter
\def\mr@ignsp#1 {\ifx\:#1\@empty\else #1\expandafter\mr@ignsp\fi}%
\newcommand{\multiref}[1]{\begingroup
\xdef\mr@no@sparg{\expandafter\mr@ignsp#1 \: }%
\def\mr@comma{}%
\@for\mr@refs:=\mr@no@sparg\do{\mr@comma\def\mr@comma{,}\ref{\mr@refs}}%
\endgroup}
\makeatother

\newcommand{\hypref}[2]{\ifx\href\asklfhas #2\else\href{#1}{#2}\fi}

\renewcommand{\eqref}[1]{(\multiref{#1})}

\ifx\href\asklfhas\newcommand{\href}[2]{#2}\fi

\newcommand{\levi}{\epsilon}

\newcommand{\deriD}{\mathcal{D}}

\begin{document}

\newpage

\DOIsuffix{theDOIsuffix}
\Volume{55}
\Month{01}
\Year{2007}
\pagespan{1}{}



\title[Free Energy of ABJM Theory]{Free Energy of ABJM Theory}


\author[M. Smedb{\"a}ck]{Mikael Smedb{\"a}ck%
  \footnote{\quad E-mail:~\textsf{mikael.smedback@physics.uu.se}
            }}
\address{Department of Physics and Astronomy, Uppsala University, Uppsala, Sweden}
\begin{abstract}
The free energy of ABJM theory has previously been computed in the strong and weak
coupling limits. In this note, we report on results for the computation of the
first non-vanishing quantum correction to the free energy, from the
field theory side. The correction can be expressed in terms of a thermal
mass for the scalar fields. This mass vanishes to 1-loop order, but there is
a non-vanishing result to 2-loop order.
Hence, the leading correction to the free energy is non-analytic in the
't Hooft coupling constant $\lambda$.
The reason is that the infrared divergences necessitate a resummation
of ring diagrams and a related reorganization of perturbation theory, in
which already the leading correction receives contributions from all orders
in $\lambda$. 
These results suggest that the free energy interpolates smoothly between
weak and strong coupling.

\end{abstract}
\maketitle                   







\setcounter{page}{1}
\renewcommand{\thefootnote}{\arabic{footnote}}
\setcounter{footnote}{0}


\setlength{\extrarowheight}{5pt}






\section{Introduction}


The gauge/gravity correspondence
\cite{Maldacena:1997re,Gubser:1998bc,Witten:1998qj}
has been studied for a long time
from both sides of the correspondence
and in various versions.
However, until recently, studies of the field theory side of the 
$AdS_4/CFT_3$ version were hampered by the fact that it seemed
difficult to write down a Langrangian with all the right symmetries,
notably superconformal
$OSp(8|4)$ symmetry and parity invariance.
In fact, many people believed such a Lagrangian was impossible to
write down, based on various no-go results.

This changed with the work of Bagger, Lambert and Gustavsson (BLG)
\cite{Bagger:2006sk,Bagger:2007jr,Bagger:2007vi,Gustavsson:2007vu}.
Bagger and Lambert wrote down an action 
in terms of an
algebraic construct known as a ``3-algebra''.
However, even though it had all the right symmetries,
closer scrutiny of the moduli space seemed to reveal
subtle difficulties in matching it to any known gravity dual,
even for $N=2$
\cite{Lambert:2008et,Distler:2008mk}.

The obstruction towards describing more than two M2-branes manifested itself in various ways in the different formulations of BLG theory. In the original 3-algebra formulation, the so-called fundamental identity only had a single unique
solution, corresponding to gauge group $SO(4)$.
In van Raamsdonk's $SU(2) \times SU(2)$ formulation
\cite{VanRaamsdonk:2008ft},
there was a reality condition on the scalars which does not make sense beyond
$N=2$. And finally, in a superspace formulation, the superpotential is only
valid for $N=2$.

This problem was finally solved in 2008, when Aharony, Bergman, Jafferis and
Maldacena introduced what is now known as the ABJM model
\cite{Aharony:2008ug}.
In this model, the superpotential was crucially rewritten in a way to allow
generalization to arbritrary ranks of the gauge group, while still
coindicing with the BLG choice for $N=2$. There are differences even for $N=2$,
since the ABJM model allows certain $U(1)$ factors,
which are not present in BLG.
This is related to the crucial role
played by monopole operators in ABJM theory, which appear to be essential
to obtaining the correct moduli space.

In this note, we will be interested in the unique behaviour of the degrees of
freedom in ABJM theory, in particular as captured by the free energy.
In super Yang-Mills theory (SYM),
the free energy has been computed at strong coupling,
as a system of D-branes \cite{Gubser:1996de,Klebanov:1996un}.
Corrections using supergravity were computed in
\cite{Gubser:1998nz,Tseytlin:1998cq}.
At weak coupling, the free energy including loop corrections was computed in 
\cite{Fotopoulos:1998es,VazquezMozo:1999ic,Kim:1999sg,Nieto:1999kc},
revealing a screening phenomenon for both scalars and the gluons,
related to the theory being strongly divergent in the infrared.

ABJM theory at finite temperature and at strong coupling was also studied recently. In 
\cite{Bak:2010yd}, a dimensional reduction of type IIA supergravity was carried out. Various static length scales including the mass gap and Debye screening mass were computed in \cite{Bak:2010qb}. An interesting phase transition which breaks the R-symmetry was found in \cite{Bak:2010ry}, and a domain wall solution was found in \cite{Yun:2010pd}.

In \cite{Smedback:2010ji}, we computed
the quantum corrections to the 
free field theory result of ABJM \cite{Aharony:2008ug},
allowing us to see whether similar phenomena as in the SYM case also
appear in ABJM. 
Our primary motivation, though, was to see if we can make progress from the field
theory side on understanding the strong suppression of the entropy at
strong coupling 
\cite{Klebanov:1996un} (corrections were computed in \cite{Garousi:2008ik}).


\section{ABJM Model}

The ABJM model
\cite{Aharony:2008ug}
is a three-dimensional Chern-Simons-matter theory
with
$\mathcal{N}=6$ superconformal symmetry.
With the gauge group $U(N) \times U(N)$ and at large $N$,
this theory is dual to M-theory on $AdS_4 \times S^7 / \mathbb{Z}_k$.

In Euclidean space, the action is
\be\label{eq_action}
  S \eq \frac{k}{2\pi} \int d^3x\: \Bigsbrk{ 
        \levi^{i j k} \tr \bigbrk{
        -\frac{i}{2} A_i \partial_j A_k + \tfrac{1}{3} A_i A_j A_k
        +\frac{i}{2} \hat{A}_i \partial_j \hat{A}_k - \tfrac{1}{3} \hat{A}_i \hat{A}_j \hat{A}_k
      }
\nl \hspace{20mm}
    + \tr (\deriD_i Y_A)^\dagger \deriD^i Y^A
    + i \tr \psi^{\dagger A} \slashed{\deriD} \psi_A
    + V^{\mathrm{bos}} + V^{\mathrm{ferm}}  
}.
\ee
The potentials are
\be
   V^{\text{bos}} \eq - \frac{1}{3} \tr \Bigsbrk{
          Y^A Y_A^\dagger Y^B Y_B^\dagger Y^C Y_C^\dagger 
      +   Y_A^\dagger Y^A Y_B^\dagger Y^B Y_C^\dagger Y^C
\nl\hspace{11mm}
      + 4 Y^A Y_B^\dagger Y^C Y_A^\dagger Y^B Y_C^\dagger 
      - 6 Y^A Y_B^\dagger Y^B Y_A^\dagger Y^C Y_C^\dagger 
   } \; , \\
   V^{\text{ferm}} \eq i \tr \Bigsbrk{
        Y_A^\dagger Y^A \psi^{\dagger B} \psi_B
      - Y^A Y_A^\dagger \psi_B \psi^{\dagger B}
      + 2 Y^A Y_B^\dagger \psi_A \psi^{\dagger B}
      - 2 Y_A^\dagger Y^B \psi^{\dagger A} \psi_B
\nl\hspace{10mm}
      - \levi^{ABCD} Y_A^\dagger \psi_B Y_C^\dagger \psi_D
      + \levi_{ABCD} Y^A \psi^{\dagger B} Y^C \psi^{\dagger D}
      } \; .
\ee
These manifestly $SU(4)$ R-symmetric potentials 
(corresponding to $\mathcal{N}=6$ supersymmetry)
are taken from \cite{Aharony:2008ug,Benna:2008zy}.
However, we have rescaled the matter so that we can factor out a factor of $k$
in front of the action.
To also see the parity-invariance, note that each of the two Chern-Simons terms
get a minus sign under parity. Hence, the sum of them is parity-invariant,
if we also interchange the role of the two gauge fields under the
parity operation.

\begin{figure}[t]
 \centering 
    \begin{picture}(0,0)
{\Large
    \put(55, 50){\makebox(0,0)[l]{$U(N)$}}
    \put(189, 50){\makebox(0,0)[l]{$U(N)$}}
    \put(-20, 50){\makebox(0,0)[l]{$A_i$}}
    \put(285, 50){\makebox(0,0)[l]{$\hat{A}_i$}}
    \put(115, 107){\makebox(0,0)[l]{$Y^A,\psi_A$}}
    \put(115, -2){\makebox(0,0)[l]{$Y^{\dagger}_A,\psi^{\dagger A}$}}
}
    \end{picture}
  \includegraphics{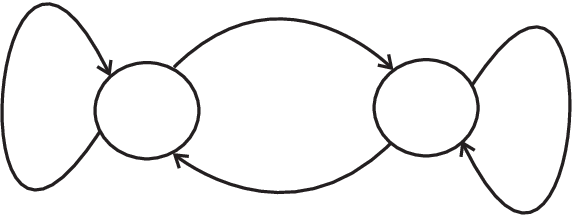}
  \caption{The ABJM model.}\label{fig_fieldcontent}
\end{figure}

The matter is in the bifundamental representation of the gauge group.
The gauge fields are in the adjoint of the left and right $U(1)$.
The field content is summarized in figure
\ref{fig_fieldcontent}.
The covariant derivative is
$
\deriD_i Y^A = \partial_i Y^A + i A_i Y^A -i Y^A \hat{A}_i.
$
The Dirac matrices are
$\left( \gamma^i \right)_\alpha^{\hspace{2mm}\beta} =\left( -\sigma^2, \sigma^1, \sigma^3 \right)$,
where $\sigma^i$ are the Pauli spin matrices.

We are interested in the thermal theory.
Hence, we compactify time, and consider the theory on
$\mathbb{R}^2 \times S^1$.
This breaks both supersymmetry and conformal invariance.
It is also important to remember that under this compactification,
fermions will be antiperiodic along the circle, corresponding
to half-integer momenta (in some unit).

We also need to gauge-fix the action (\ref{eq_action}).
Possible choices include Lorentz gauge and Coulomb gauge.
Lorentz gauge makes the combinatorics simpler.
On the other hand, Coulomb gauge is often very convenient
in thermal calculations. In particular, Coulomb gauge makes
gauge fields and ghosts manifestly non-propagating.
In this note, we will use Coulomb gauge.


\section{Infrared divergences}

We now want to investigate the coupling dependence of the free energy.
As discussed in \cite{Smedback:2010ji}, one of the technical difficulties is
the infrared divergences.
%
%
%
Assuming that the free energy is analytic in $\lambda=N/k$
around $\lambda=0$, we write
\begin{equation}\label{eq_perturbativeF}
  \tilde{F}(\lambda)=\tilde{F}_1 + \lambda \tilde{F}_2 + \lambda^2 \tilde{F}_3 + \cdots.
\end{equation}
To one-loop order,
only scalars and fermions contribute.
Summing up those contributions, we find
\begin{equation}\label{eq_Fabjm}
  \tilde{F}_1 = N^2 \left(8 \frac{1}{2}A_0-8\frac{1}{2}A_{1/2}\right) =
 -N^2 T^3 \frac{7 \zeta({3})}{\pi},
\end{equation}
where
\begin{equation}\label{eq_A}
\begin{split}
  A_{\nu} & = \int \frac{d^2 p}{(2\pi)^2} T \sum_{n= \in \mathbb{Z}+\nu}  \log ( \vec{p}^2+\omega_n^2 ), \\
 A_{0}   & =-T^3 \frac{\zeta(3)}{\pi}, \hspace{10mm}
  A_{1/2} =T^3 \frac{3\zeta(3)}{4 \pi}.
\end{split}
\end{equation}
The reason that gauge fields and ghosts do not contribute is that the gauge
field does not have proper kinetic terms, only Chern-Simons terms.
At zero temperature, supersymmetry ensures that the free energy cancels out,
and this is also true of the expression (\ref{eq_Fabjm}).

The result (\ref{eq_Fabjm}) was already obtained in
\cite{Aharony:2008ug}.
To attempt to find further terms in the naive expansion (\ref{eq_perturbativeF}),
we use perturbation theory.
Propagators and vertices follow from the action
(\ref{eq_action}).
At each order, connected and one-particle-irreducible diagrams
contribute.
At two loops, all such diagrams vanish, so we find
that $\tilde{F}_2=0$.
Hence, it would appear we need to proceed to three-loop order.


However, the perturbative expansion
(\ref{eq_perturbativeF}) breaks down somewhere between two and
three loops, due to infrared divergences.
For example, the 6-vertex will contribute a correction proportional
to 
\begin{equation}
  \left(  \int \frac{d^2 p}{(2\pi)^2}   T \sum_{n \in \mathbb{Z}} 
    \frac{1}{\vec{p}^2+\omega_n^2} \right)^3.
\end{equation}
These infrared divergences do not cancel.


\begin{figure}[t]
{\LARGE $\sum_{N}$} 
  \centering
 \raisebox{-25mm}{
    \begin{fmffile}{diagring}
    \begin{fmfgraph*}(150,150)
       \fmfleft{i1,i2}
       \fmfright{o1}
       \fmfv{decor.shape=circle,decor.filled=empty}{v1}
       \fmfv{decor.shape=circle,decor.filled=empty}{v2}
       \fmfv{decor.shape=circle,decor.filled=empty}{v3}
       \fmflabel{$1$ \hspace{0mm}}{v1}
       \fmflabel{$2$ \hspace{0mm}}{v2}
       \fmflabel{\hspace{0mm} $N$}{v3}
       \fmf{phantom}{i1,v1}
       \fmf{phantom}{i2,v2}
       \fmf{phantom}{o1,v3}
       \fmf{plain,tension=0.2,left=0.6}{v1,v2}
       \fmf{dots,tension=0.2,left=0.6}{v2,v3}
       \fmf{plain,tension=0.2,left=0.6}{v3,v1}
    \end{fmfgraph*}
    \end{fmffile}}
\caption{Free energy ring diagrams.}
\label{fig_ring}
\end{figure}

\section{Thermal Mass}\label{sec_thermal}

Since the perturbative expansion
(\ref{eq_perturbativeF})
breaks down, we need to reorganize pertubation theory
to obtain a finite answer beyond two loops.
Equivalently, we regularize the theory by introducing
a thermal mass.
How this works was already studied both
in QCD
\cite{Arnold:1994ps,Arnold:1994eb,Kapusta:1979fh,Gross:1980br}
and in super Yang-Mills
\cite{VazquezMozo:1999ic,Kim:1999sg,Nieto:1999kc}.


Specifically, the idea is that we write the action
$S = \left( S + \delta S_2 \right) - \delta S_2$,
where (in momentum space)
\begin{equation}\label{eq_deltaS2}
\begin{split}
  \delta S_{\text{2}}  = \frac{k}{2\pi} \int 
  \frac{d^2p}{(2\pi)^2} \; T \sum_{n \in \mathbb{Z}}  \tr \text{{\Huge [}} 
& \frac{1}{2}
  Y^\dagger_A(p) m^2_{Y} Y^A(-p) \text{{\Huge ]}} .
\end{split}
\end{equation}
Treating $- \delta S_2$ as a perturbation to $\left( S + \delta S_2 \right)$
creates a thermal mass in the scalar propagator.
A thermal mass can be interpreted as the appearance of
a screening length
$r=\frac{1}{m}$
\cite{Gross:1980br}.

We compute the scalar self-energy by summing up
all one-particle-irreducible diagrams.
It will be sufficient for our purposes to do this in the static limit, 
corresponding to
no external momentum, since this is the troublesome mode
that we want to regularize.

To two-loop order,
we find the result \cite{Smedback:2010ji}
\begin{equation}\label{eq_scalar_summary}
\begin{split}
  m_Y^2(\lambda) & = (2\pi T)^2\mu^2(\lambda), \\
  \mu^2(\lambda) 
& = \frac{118}{3 (2\pi)^2} 
  \lambda^2 \log(\mu)^2 
  + 
  \mathcal{O}(\lambda^2 \log(\lambda)),
\end{split}
\end{equation}
corresponding to a regularized scalar propagator
\begin{equation}\label{eq_rens}
 \raisebox{-12mm}{
    \begin{fmffile}{diagselfenergyYY}
    \begin{fmfgraph}(75,75)
       \fmfleft{i1}
       \fmfright{o1}
       \fmfblob{.30w}{v1}
       \fmf{plain}{o1,v1,i1}
       \fmfdot{i1,o1}
    \end{fmfgraph}
    \end{fmffile}}
=
  \frac{2\pi}{k T} \, \frac{1}{\vec{p}^2+\omega_n^2+m_Y^2}.
\end{equation}

Note that the computation
needs to be done self-consistently by requiring that successive
higher order perturbative corrections
do not shift the pole in the propagator (cf. D'Hoker's original calculation for $\text{QCD}_3$ \cite{D'Hoker:1981us}).
Hence, (\ref{eq_scalar_summary}) is an equation
which in general cannot be explicitly solved for $m_Y^2$.


\begin{figure}[t]
  \centering
  \subfigure[Free energy.]{\includegraphics[height=4cm]{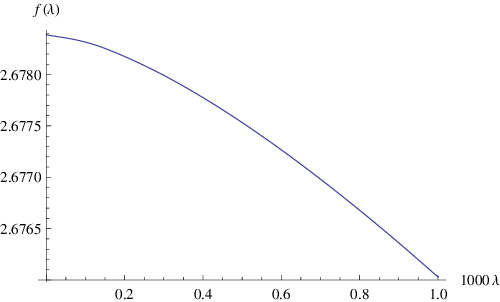}\label{fig_free}}
  \qquad
  \subfigure[Scalar thermal mass.]{\includegraphics[height=4cm]{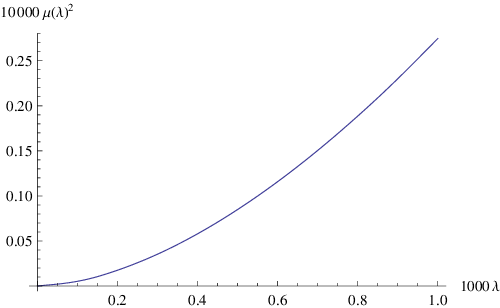}\label{fig_scalar}}
  \caption{Free energy and scalar thermal mass.}\label{fig_fs}
\end{figure}


\section{Reorganized Perturbation Expansion}

We have seen that due to IR divergences, the naive expansion
 (\ref{eq_perturbativeF}) is not successful in finding quantum corrections
to the free energy.
Instead, we will have to use the reorganized perturbation expansion
\begin{equation}\label{eq_Ftotal}
  F(\lambda)=F_1(\lambda) + \lambda F_2(\lambda) + \lambda^2 F_3(\lambda) \cdots,
\end{equation}
where the coefficients $F_i(\lambda)$ now depend on $\lambda$.
The reason is that we are now using the massive scalar propagator
(\ref{eq_rens}).
%
%
%
At one-loop order, we find
\begin{equation}\label{eq_F1resummed}
  F_1(\lambda) = N^2 \left(8 \frac{1}{2}A_0(m_Y^2)-8\frac{1}{2}A_{1/2}(m_{\Psi}^2)\right),
\end{equation}
which generalizes the zero-mass answer (\ref{eq_Fabjm}).
However, from the derivation in section \ref{sec_thermal},
it follows that already the one-loop answer is sensitive to corrections
to all orders in $\lambda$.
To obtain the first non-vanishing correction, it will be sufficient
to know the contribution from the scalars
(see figure \ref{fig_ring}),
\begin{align}\label{eq_Aperiodic}
  A_0(m^2) & = T \sum_{n\in\mathbb{Z}}  \int \frac{\text{d}^2 p}{(2 \pi)^2} \log{\left( \vec{p}^2+ \left(2\pi T n\right)^2+m^2\right)} = \\ \nonumber
& =\frac{T^3}{4\pi} \left[ -4 \zeta(3) - m^2 T^{-2} \log{(m^2T^{-2})}  
+m^2T^{-2}
+\mathcal{O}\left(m^4T^{-4}\right)
 \right],
\end{align}
which generalizes (\ref{eq_A}).
No thermal mass is generated for the fermions, since they have no zero modes.
Hence, they only contribute to lower orders.
%
%
%
In this way, we obtain the free energy including the first non-vanishing
quantum correction, as already reported on in \cite{Smedback:2010ji}.
Specifically, by combining equations
(\ref{eq_Ftotal}),
(\ref{eq_F1resummed})
and
(\ref{eq_Aperiodic}),
we find the free energy density\footnote{
Using the approximation $\log(\mu)\approx \log(\lambda)$ 
(valid for very small $\lambda$),
it can be confirmed that to lowest non-vanishing order, the $\lambda$
dependence of the correction is $\lambda^2 \log(\lambda)^3$, i.e. the same as that found by Gaiotto and Yin for 
a related type of three-dimensional Chern-Simons-matter theory 
\cite{Gaiotto:2007qi}.
}
\begin{equation}\label{eq_free_final}
  F=-N^2 T^3 f(\lambda),
\end{equation}
where
\begin{equation}\label{eq_f}
  f(\lambda)=\left[
    \frac{7 \zeta(3)}{\pi}
    + \frac{m_Y^2(\lambda)}{\pi T^{2}} 
    \log\left(\frac{m_Y^2(\lambda)}{ T^{2}} \right)
    + \mathcal{O}(\lambda^2 \log(\lambda)^2)
  \right].
\end{equation}


\noindent
The scalar thermal mass (\ref{eq_scalar_summary})
can be solved for numerically.
It and the free energy (\ref{eq_f}) are plotted in
figure \ref{fig_fs}. 
The result is suggestive of a free energy which smoothly interpolates to
the strong coupling answer \cite{Klebanov:1996un}
\begin{equation}
  f_{\text{s}}(\lambda) = \left[ \frac{2^{7/2}}{9} \pi^2 \frac{1}{\sqrt{\lambda}} + \cdots \right],
\end{equation}
as expected.


\section{Conclusions}

In this note, we described how to calculate the free energy
of ABJM theory on $\mathbb{R}^2 \times S^1$.
The first non-vanishing quantum correction is non-analytic
in $\lambda$ \cite{Smedback:2010ji}. 
The reason is that perturbation theory had to
be reorganized to deal with the IR divergences (resummation
of ring diagrams). Thus, the theory 
was regularized by a scalar thermal mass, generated by
screening effects.
This calculation suggests that the free energy interpolates 
smoothly between weak and strong coupling.


\begin{acknowledgement}

I thank Lisa Freyhult, Joseph Minahan, Olof Ohlsson Sax, Diego Rodriguez-Gomez, Edward Witten, Amos Yarom and in particular Igor Klebanov, Thomas Klose and Juan Maldacena for several very useful discussions. I thank Igor Klebanov and Thomas Klose for initial collaboration on the work reported on in this note.
This work was done while I was affiliated with Princeton University
and Uppsala University.
This research was supported by
a Marie Curie Outgoing International Fellowship, contract No. MOIF-CT-2006-040369, within the 6th European Community Framework Programme.

\end{acknowledgement}











\bibliographystyle{nb}
\bibliography{entropy}

\end{document}